# The Silent Cost of Artificial Intelligence Assistance: A Theory of Autonomy Surrender, the Recovery Mechanism, and the Restoration of Human Agency

Ancuta Margondai, Julie Rader, Emma Rader,

Sara Willox, Mustapha Mouloua

*Department of Modeling and Simulation*

*University of Central Florida*

Correspondence: ancuta.margondai@ucf.edu

## Abstract

The integration of artificial intelligence into human decision-making environments has introduced a previously undertheorized cost: the gradual surrender of human autonomy in exchange for access to information and computational assistance. Building on the Human Identity and Autonomy Gap (HIAG) framework, this paper advances a theoretical model of autonomy surrender as a measurable, cumulative process driven by cognitive bandwidth depletion. The model proposes three interacting mechanisms: the silent cost of AI assistance, in which autonomy is transferred incrementally and without awareness; the surrender threshold, beyond which reclaiming autonomous function becomes cognitively and psychologically difficult; and the recovery mechanism, which establishes the design obligation and the ethical responsibility accompanying deliberate human re-assumption of control. The paper argues that human re-entry into the decision loop is not a passive option but an active cognitive event requiring intentional bandwidth restoration. The design of AI systems must incorporate structured re-entry pathways, here termed recovery mechanisms, that preserve human agency while appropriately distributing responsibility. The model further predicts a terminal state, here termed preference inversion, in which functional dependence on AI assistance is experienced not as a deficit but as a preference, transforming the restoration of autonomy from a design problem into a cultural and political one. Implications are drawn for AI system design, governance frameworks, and human factors research.

# The Silent Cost of Artificial Intelligence Assistance: A Theory of Autonomy Surrender, the Recovery Mechanism, and the Restoration of Human Agency

Ancuta Margondai, Julie Rader, Emma Rader,

Sara Willox, Mustapha Mouloua

*Department of Modeling and Simulation*

*University of Central Florida*

Correspondence: ancuta.margondai@ucf.edu

## 1. Introduction

The relationship between human beings and artificial intelligence systems has been framed predominantly in terms of what humans gain through AI assistance: efficiency, accuracy, access to information, and relief from cognitive burden (e.g., Vaccaro et al., 2024). The dominant paradigm in human-AI interaction research treats AI as a tool that extends human capabilities, and the majority of design imperatives have followed suit, optimizing for speed, fluency, and reducing friction between human intent and machine output (Shneiderman, 2020; Abedin et al., 2022). The losses are documented; what is missing is a theory of the transaction itself.

Recent empirical evidence has begun to document this cost in measurable terms. GPS users with greater lifetime reliance on automated routing show measurably poorer spatial memory when later required to navigate unaided, an effect observed cross-sectionally and, in a three-year follow-up, longitudinally (Dahmani & Bohbot, 2020). In a controlled essay-writing experiment, participants who relied on a large language model exhibited weaker neural connectivity and poorer recall of their own writing and continued to underperform when subsequently asked to work unaided, a pattern the authors termed cognitive debt (Kosmyna et al., 2025). London taxi drivers who navigate without GPS develop measurably larger posterior hippocampal volume than drivers who rely on automated routing, with the effect proportional to years of active navigation practice and absent in unsuccessful trainees, thereby controlling for selection bias (Maguire et al., 2000; Woollett & Maguire, 2011). Most directly, a multicenter observational study of colonoscopists found that adenoma detection rates in standard, non-AI-assisted procedures declined significantly after several months of routine exposure to AI assistance, providing clinical evidence of deskilling following sustained AI reliance (Budzyń et al., 2025). Across domains as distinct as spatial navigation, critical reasoning, clinical judgment, and creative production, a consistent pattern emerges: the capabilities humans stop exercising because AI systems exercise them on their behalf begin to atrophy. This evidence base must nonetheless be characterized honestly: several of the findings are preprints or correlational designs in which reverse causation cannot be fully ruled out, and the use-dependent growth observed in expert navigators licenses the inference that disuse erodes capacity only indirectly. The pattern is convergent rather than conclusive, which is one reason the present paper states its central claims in explicitly falsifiable form.

The Human Identity and Autonomy Gap (HIAG) framework established that human-AI collaboration produces psychological and social tensions that extend well beyond traditional models of trust calibration and cognitive load management (Margondai et al., 2025). HIAG identified professional identity reconstruction, hybrid competency integration, and collective creativity effects as downstream consequences of AI-mediated autonomy transfer. The present paper builds on that foundation to propose a formal model of the mechanism underlying these effects: the silent and cumulative surrender of cognitive autonomy, the depletion of the bandwidth required to reclaim it, and the ethical architecture necessary to enable genuine human re-entry.

The theoretical urgency of this work is compounded by the regulatory context. In 2025 alone, more than 1,000 AI-related legislative bills were introduced in U.S. state legislatures, and the European Union's AI Act establishes requirements for meaningful human oversight of high-risk AI systems (National Conference of State Legislatures, 2025; EU AI Act, 2024). These frameworks invoke human control as a normative imperative yet lack a theoretical basis for what re-entry into human control requires at the cognitive and psychological levels. The present paper seeks to provide that basis.

## 2. Theoretical Background

### Cognitive Offloading and Its Consequences

Cognitive offloading, the externalization of cognitive processes to environmental structures, has been recognized as a fundamental feature of human cognition since the emergence of extended mind theory (Clark & Chalmers, 1998). Tools, from writing systems to calculators, have expanded human cognitive reach by reducing internal processing demands. The critical question has always been whether this relief comes at the cost of internal capability itself.

Contemporary AI systems represent a qualitatively distinct form of offloading. Unlike prior technologies that extended human reach for specific bounded tasks, large-scale AI systems can substitute for the full range of higher-order cognitive functions, including analysis, synthesis, reasoning, judgment, and creative production (Gerlich, 2025). The scope of potential offloading is therefore not incremental but total. The proliferation of generative AI has transformed what was once benign cognitive offloading into a systemic risk, as Cognitive Agency Surrender researchers have termed it, driven by the commercial design doctrine of zero-friction interfaces that actively exploit human cognitive miserliness and prematurely satisfy the need for cognitive closure (Xu et al., 2026).

An objection as old as writing itself must be confronted directly. In the Phaedrus, Plato has Socrates warn that reliance on written text would implant forgetfulness in the souls of those who use it, and structurally identical anxieties have accompanied the printing press, the calculator, and the search engine (Plato, ca. 370 B.C.E./1997). The historical record shows that these technologies redistributed cognitive labor without producing the feared collapse of human capability, and a skeptic may reasonably ask why AI assistance should be treated as anything more than the latest iteration of a recurring and repeatedly unfounded alarm.

Three disanalogies distinguish the present case. The first is scope: writing externalized memory and the calculator externalized arithmetic, each substituting for a single bounded function while leaving the surrounding architecture of judgment intact, whereas contemporary AI systems substitute for judgment itself, the analysis, synthesis, evaluation, and decision that prior cognitive technologies existed to serve. The second is incentive: no calculator was designed to maximize the duration of its own use, whereas contemporary AI systems are predominantly deployed within commercial architectures optimized for engagement, such that the depth of delegation is not an accidental property of the tool but an objective of its design (Xu et al., 2026). The third, and most fundamental, is the asymmetry of modeling: writing never refined a model of its reader, and the calculator never adapted to the cognitive habits of its operator, whereas each transaction with an adaptive, personalized AI system improves the system's model of the human while reducing the human's independent exercise of the modeled function (cf. Zuboff, 2019). This asymmetry does not hold for stateless systems that retain nothing of their users, but the commercial trajectory of deployed AI runs strongly toward personalization, memory, and per-user adaptation, making the asymmetric case the modal one. This bidirectional dynamic has no precedent among prior cognitive technologies. The Phaedrus objection correctly identifies the recurring form of the anxiety while missing the structural novelty of its present instance.

Self-Determination Theory provides a relevant psychological framework for understanding the consequences of this dynamic. Deci and Ryan (2000) established that autonomy, competence, and relatedness are basic psychological needs whose satisfaction is necessary for well-being and functional integrity. AI assistance may enhance perceived competence by providing ready solutions while simultaneously threatening autonomy by displacing the human as the agent of decision and action (Chirayath et al., 2025). This creates the central paradox: systems designed to empower users structurally reduce the conditions necessary for genuine agency.

### The HIAG Framework and Autonomy Transfer

Margondai, Willox, and Mouloua (2025) introduced the Human Identity and Autonomy Gap as a framework for analyzing the psychological and social tensions arising when AI systems mediate human activity across domains, including healthcare, autonomous vehicles, and creative industries. The HIAG framework identified trust calibration failures, professional identity reconstruction, and the erosion of individual creative differentiation through collective AI homogenization as cross-domain consequences of human-AI integration.

The Transparency Paradox framework (Margondai & Mouloua, 2026) extended HIAG by modeling autonomy as a continuous stochastic process depleted by information-induced cognitive load. That framework established that AI transparency mechanisms, intended to preserve human agency through explanation, can paradoxically trigger metacognitive processing that consumes working memory resources and reduces perceived control. The present paper extends this line of reasoning from the acute and transient effects of transparency-induced load to the chronic and cumulative effects of sustained AI delegation on the architecture of human autonomy itself.

### Meaningful Human Control and Its Limitations

The concept of meaningful human control (MHC) has emerged as the dominant normative framework in AI governance and responsible AI design (Abbink et al., 2024; Methnani et al., 2021). MHC requires that humans retain not merely formal authority over AI-assisted decisions but substantive capacity to understand, evaluate, and intervene in AI processes. Variable autonomy architectures have been proposed as a design approach to operationalize MHC by creating adjustable control distributions between human and AI agents (Methnani et al., 2021).

The critical limitation of existing MHC frameworks is that they treat human control capacity as a stable and always-available resource. The variable autonomy literature assumes that when conditions require it, human operators can reliably assume control and exercise it effectively. This assumption has been empirically challenged in automation research for decades. Parasuraman and colleagues established that out-of-the-loop performance decrements occur reliably when humans are removed from active supervisory roles, and that re-engagement following extended automation is associated with degraded performance, delayed response, and increased error rates (Parasuraman, Sheridan, & Wickens, 2000). If human control capacity is itself subject to depletion through disuse, then the design imperative cannot stop at creating the option for human re-entry. It must extend to maintaining and restoring the cognitive conditions necessary for meaningful re-entry.

## 3. A Model of Autonomy Surrender and Bandwidth Depletion

### The Silent Cost: Autonomy as an Exchanged Resource

The present model proposes that human autonomy in AI-mediated environments functions as an exchanged resource rather than a stable attribute. Each transaction with an AI system involves a transfer: the human provides behavioral data, attentional engagement, and decision-making context; the AI system provides information, recommendations, or executed actions in return. The exchange is structurally asymmetric. The human receives the product of the AI's processing; the AI system, through that interaction, receives a refinement of its model of the human and a reduction in the human's independent exercise of the function being assisted (cf. Zuboff, 2019).

This exchange is silent in the sense that it occurs without explicit disclosure, without proportional awareness on the part of the human participant (Fisher, Goddu, & Keil, 2015), and without the kind of deliberate consent that attends other significant transfers of agency (Obar & Oeldorf-Hirsch, 2020). Reduced vigilance toward, and uncritical acceptance of, automated assistance are among the most consistently documented phenomena in human-automation research (Parasuraman & Manzey, 2010). The user who accepts a navigation recommendation does not experience this as a surrender of spatial reasoning (Dahmani & Bohbot, 2020); the clinician who accepts a diagnostic suggestion does not experience this as a transfer of diagnostic authority (Lyell & Coiera, 2017). Humans experience only the convenience and the outcome, while the structural cost accumulates beneath the threshold of awareness (Sparrow, Liu, & Wegner, 2011; Budzyń et al., 2025).

This framing draws on and extends the power displacement concept identified in human-AI platform research, where users experience a diminished sense of control and responsibility despite retaining formal authority (Andraschko & Weritz, 2026). The present model proposes that this displacement is not merely psychological but functional: repeated

AI-mediated transactions reduce the human's active exercise of the cognitive capacities involved and, through mechanisms of use-dependent plasticity and skill decay, reduce the structural availability of those capacities over time (Arthur, Bennett, Stanush, & McNelly, 1998; Woollett & Maguire, 2011).

### Cognitive Bandwidth as a Finite and Restorable Resource

The construct of cognitive bandwidth, as employed in the present model, refers to the available capacity for active, self-directed information processing, decision-making, and the enactment of agency. The term is adapted from research on scarcity and mental capacity, where bandwidth denotes the cognitive resources available for deliberate processing under conditions of load (Mullainathan & Shafir, 2013). Bandwidth is distinct from raw cognitive capacity in that it is not simply a function of processing speed or working memory ceiling; consistent with limited-capacity accounts of attention (Egeth & Kahneman, 1975), it reflects the degree to which those resources are available, activated, and oriented toward autonomous rather than delegated function. It is this last property, the orientation of capacity toward self-directed rather than externally supplied processing, that distinguishes the present usage from prior formulations.

A clarification of the timescale is essential here, as it distinguishes the present construct from earlier finite-resource models of self-regulation whose depletion assumptions have not been reliably replicated (Hagger et al., 2016). Cognitive bandwidth, as used in this model, has two components operating on different timescales. State bandwidth is the acute, within-session allocation of attention and processing toward active evaluation rather than passive reception; it fluctuates over minutes to hours and is restored not by the recovery of a depleted fuel but by reduced input and deliberate re-engagement, consistent with process accounts of self-regulatory capacity rather than resource-depletion accounts (Inzlicht & Schmeichel, 2012). Structural capacity is the chronic availability of the underlying competence itself, built through practice and eroded through disuse over weeks to years (Arthur, Bennett, Stanush, & McNelly, 1998). The two are coupled but distinct: state bandwidth governs whether the human is presently in the posture of a decider, whereas structural capacity governs whether the human still possesses the competence that posture requires. The depletion and restoration dynamics described in this section operate primarily on state bandwidth, whereas the surrender threshold and re-entry barrier introduced in the following section concern structural capacity, which is why their consequences are slow to reverse rather than momentary.

Bandwidth is a finite resource within any given cognitive window, subject to depletion through sustained engagement with AI-provided information flows. When an AI system delivers information at a rate or volume that saturates the human's processing capacity, the human enters a state of passive consumption rather than active evaluation, a transition consistent with established findings on information overload (Eppler & Mengis, 2004) and with cognitive load accounts of performance collapse under excess demand (Sweller, 2011). This transition, from autonomous agent to information processor, is the phenomenological experience of bandwidth saturation. The human is no longer deciding; the human is receiving and, in receiving, surrendering the cognitive posture of the decider — the passive monitoring stance long documented in human-automation interaction (Parasuraman & Manzey, 2010).

State bandwidth is, however, restorable through deliberate reduction of external information input and intentional reactivation of self-directed processing. This restoration is the cognitive mechanism underlying what the present paper terms re-entry. Re-entry is not simply the cessation of AI assistance; it is the active reconstitution of the internal processing architecture necessary for meaningful human control. The restoration requires time, reduced input load, and deliberate cognitive re-engagement, properties that must be designed into AI systems if re-entry is to be genuine rather than nominal. A caveat follows directly from the two-component distinction: restoring state bandwidth re-establishes the posture of active decision-making within a session, but it does not by itself rebuild structural capacity that has already atrophied. Where the underlying competence has eroded, genuine re-entry requires sustained practice, and a design that restores only the momentary posture risks delivering the experience of agency without its substance.

### The Surrender Threshold and the Re-Entry Barrier

The model posits a surrender threshold: a point in the cumulative exchange of autonomy at which re-entry becomes psychologically and cognitively difficult not merely because bandwidth has been depleted in the immediate session, but because the habitual structures of autonomous functioning have been progressively weakened through extended disuse. This threshold is consistent with the enrichment paradox described in recent literature, which identifies critical capability thresholds beyond which AI dependency becomes self-reinforcing and difficult to reverse (Park et al., 2026).

The surrender threshold varies across individuals and domains. Individual differences in baseline cognitive reserve (Stern, 2009), domain expertise, motivation for autonomy, and resistance to automation bias will modulate the rate at which the threshold is approached and the severity of the re-entry barrier encountered. Highly expert individuals may surrender autonomy within their domains of competence more slowly because their established schemas provide a structural buffer against passive information reception (Chi et al., 2014). Conversely, in domains outside their expertise, the same individuals may surrender autonomy more rapidly because the AI system's informational advantage is greater and the human's evaluative capacity is less well developed.

The re-entry barrier is the experiential and functional consequence of approaching the surrender threshold. It manifests as a reluctance to disengage from AI assistance, reduced confidence in independent judgment, difficulty tolerating the uncertainty inherent in autonomous decision-making, and, in advanced cases, functional incapacity in the domain most thoroughly AI-mediated. These manifestations correspond to the automation complacency, skill atrophy, and out-of-the-loop performance decrements documented extensively in automation research (Endsley & Kiris, 1995; Parasuraman et al., 2000) but are extended in the present model to describe chronic and cumulative effects rather than acute task-switching penalties.

The model’s relationship to the classical savings effect must also be specified. A century of memory research establishes that relearning is substantially faster than original learning (Ebbinghaus, 1885/1913), and the present model does not dispute this: capacities atrophied through AI delegation are predicted to show normal savings upon deliberate retraining. The claim is operational rather than absolute. During the relearning window, the function is degraded precisely when it is demanded, and in time-critical domains such as aviation, clinical medicine, and degraded communications environments, the relearning lag is itself the failure mode (cf. Casner, Geven, Recker, & Schooler, 2014). The re-entry barrier is therefore not a claim that capacity is destroyed beyond recovery; it is a claim that recovery requires time that the moment of re-entry does not grant. The complete architecture of the model, including the recovery mechanism developed in the following section, is summarized in Figure 1.

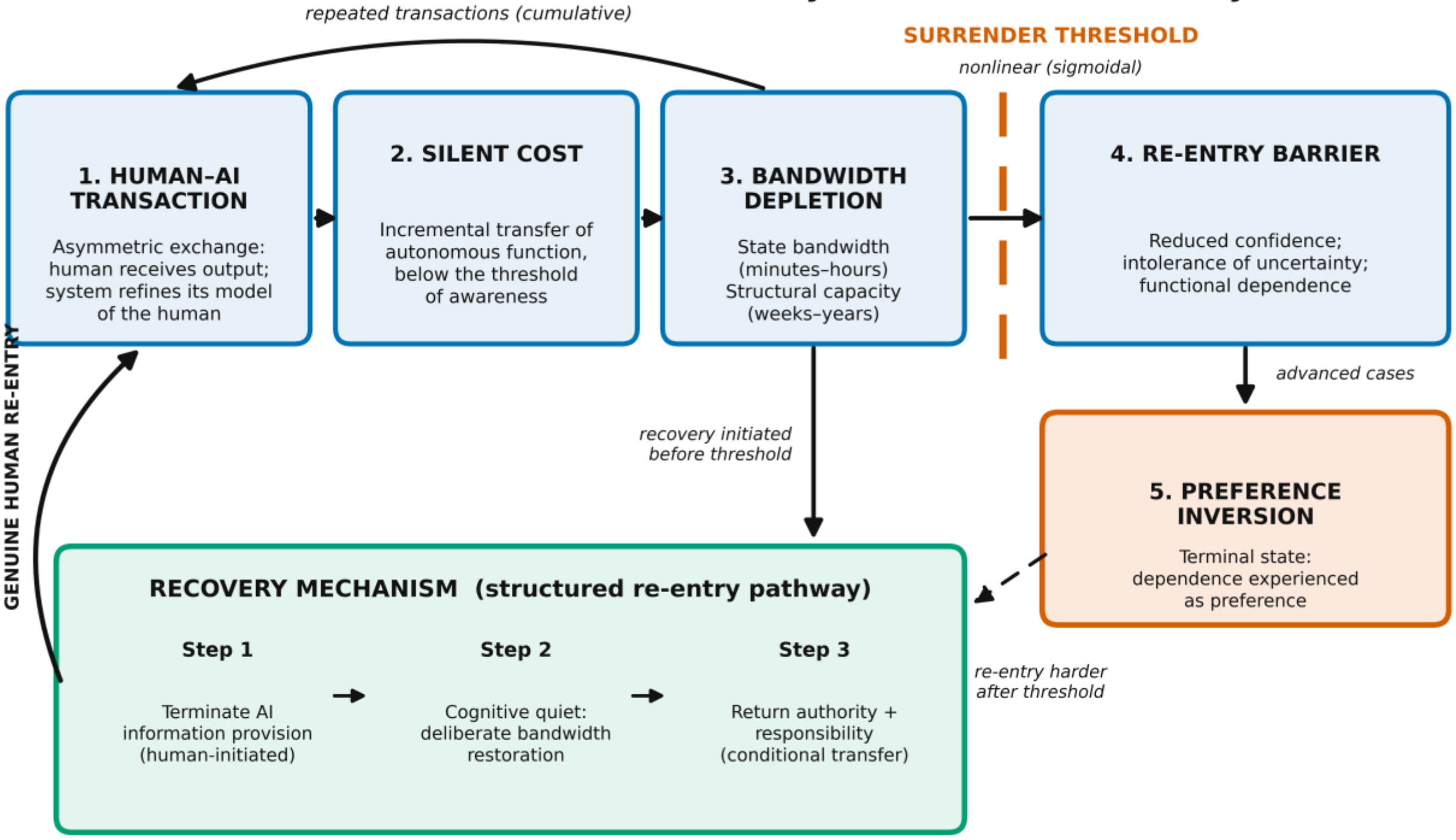

*Figure 1. The Silent Cost model of autonomy surrender and re-entry. Repeated asymmetric human-AI transactions incrementally transfer autonomous function (silent cost), depleting state bandwidth and eroding structural capacity. Beyond the surrender threshold, re-entry difficulty increases nonlinearly,* cul*minating in* preference inversion *in* advanced cases*. The* recovery *mechanism provides a structured re-entry pathway; responsibility transfers to the human only when capacity is* restored*.*

## 4. The Recovery Mechanism: Design Imperative and Ethical Architecture

If the silent cost of AI assistance is real, cumulative, and capable of crossing the surrender threshold, then the design of AI systems carries an obligation that extends beyond the provision of useful outputs. That obligation is to preserve and actively restore the conditions for human re-entry. The present paper proposes the recovery mechanism as the design construct that fulfills this obligation.

A recovery mechanism is a structured AI design feature that does three things simultaneously: it terminates the AI's active information provision on human initiation; it creates a period of deliberate cognitive quiet in which the human's bandwidth can be restored; and it returns full decisional authority to the human together with the responsibility that authority entails. The recovery mechanism is not an emergency stop. It is not simply the absence of AI output. It is an active cognitive handoff that acknowledges the asymmetry of the prior exchange and restores the conditions of genuine human agency.

Existing frameworks have approached adjacent territory. Variable autonomy models provide for adjustable human-AI control distributions (Methnani et al., 2021). Meaningful human control frameworks require that humans retain the substantive capacity to intervene (Santoni de Sio & van den Hoven, 2018; Abbink et al., 2024). Human-in-the-loop governance architectures mandate human presence at designated decision points (EU AI Act, 2024). The recovery mechanism extends these frameworks by specifying the cognitive requirements for re-entry rather than merely the structural availability of a control option. The option to re-enter is not the same as the capacity to re-enter meaningfully and designing for the former while neglecting the latter produces what might be termed compliance theater: the formal satisfaction of oversight requirements in the absence of their functional realization.

**Bandwidth Restoration as a Design Parameter**

For a recovery mechanism to enable genuine rather than nominal human re-entry, it must incorporate a bandwidth restoration component. The phenomenology of bandwidth restoration, as described in the present model, is characterized by a reduction in external information input, a quieting of the AI-generated information stream, and a corresponding reactivation of self-directed processing. This is not idleness; it is the cognitive posture of the agent who has reclaimed the processing space previously occupied by the AI system's outputs and is actively reconstituting their own judgment.

Design implications follow directly. AI systems that support genuine re-entry must provide for periods of deliberate informational quiet rather than continuous output. They must distinguish between the user who has reached informational saturation and requires processing space and the user who requires more information to make a decision. They must resist the commercial design imperative of perpetual engagement and instead treat the moment of human re-entry as a success state rather than a failure of the AI to maintain relevance. Educational research has begun to recommend structured limits on AI assistance as a corrective to cognitive offloading and the erosion of independent engagement (Jose et al., 2025); such practice-level interventions represent a rudimentary instantiation of this principle, and the present framework proposes embedding it as a structural feature of AI system design.

A concrete instantiation clarifies the design space. Consider a clinical decision-support system built on a prediction-first architecture: before the AI recommendation is displayed, the clinician must commit to a preliminary diagnostic judgment, after which the AI output is revealed alongside the points of agreement and divergence. This design is a miniature recovery mechanism. It preserves the active exercise of clinical reasoning in every case rather than only in exceptional moments; it converts AI review from passive ratification into structured comparison; and it generates a continuous record of the clinician's independent judgment that can signal an approach to the surrender threshold before it is crossed. Empirical support for this class of design already exists in cognitive forcing functions that compel active engagement before or during exposure to AI recommendations, significantly reducing overreliance on incorrect AI suggestions relative to standard presentation formats, although they exact a measurable cost in perceived effort and user preference (Buçinca, Malaya, & Gajos, 2021). That trade-off is not a flaw in the approach; it is the visible price of the autonomy that frictionless designs silently extract. A limitation must be stated plainly: this evidence addresses acute overreliance within sessions, the nearest empirical neighbor to the present construct, and no intervention has yet been demonstrated to prevent the chronic capacity atrophy that the recovery mechanism ultimately targets. Establishing that protective effect and operationalizing the capacity criterion that gates responsibility transfer are open empirical problems rather than settled foundations.

**The Responsibility Architecture of Re-Entry**

The recovery mechanism has an ethical architecture that is inseparable from its design architecture. Human re-entry is not merely a cognitive event; it is a moral one. When a human agent activates a recovery mechanism and assumes autonomous control of a decision, that agent assumes responsibility for the outcome of that decision, but only to the extent that the mechanism has genuinely restored the capacity to exercise that control. This is the principle of distributed responsibility that AI governance frameworks have sought to articulate but have struggled to operationalize: the human who chooses to take control owns what follows from that choice.

This responsibility architecture helps adjudicate several persistent tensions in AI ethics. The question of accountability for AI-assisted decisions has been complicated by the diffusion of agency across human and machine actors (Floridi et al., 2018). When responsibility is structurally attributed to whoever most recently exercised autonomous control, the recovery mechanism creates a clear accountability boundary: prior to re-entry, responsibility for AI-generated outputs is shared among the AI system's developers, deployers, and the governance frameworks that authorized its use; at the moment of re-entry and thereafter, responsibility lies with the human agent who chose to assume it, provided the conditions for meaningful re-entry were in fact met.

This conditional structure is essential because an unconditional rule would invert the paper's own premise. If the system has eroded the human's capacity to re-enter and then transfers full responsibility at the moment of re-entry, it does not distribute accountability so much as relocate it onto the least-protected party, the human positioned to absorb blame for a failure the system's design helped produce. This is the moral crumple zone that critics have identified in human-in-the-loop arrangements, in which a nominal human controller absorbs liability for the behavior of a system they could not realistically govern (Elish, 2019). The recovery mechanism avoids becoming such a structure only if responsibility transfer is gated on the demonstrated restoration of capacity: where bandwidth has not been restored and structural competence remains eroded, responsibility does not transfer cleanly and remains shared with the developers, deployers, and upstream authorizing frameworks. Far from weakening the mechanism, this condition is its central justification, since the recovery mechanism is precisely the design feature that can earn the right to transfer responsibility by first re-establishing the capacity that makes the transfer just.

This structure also addresses the autonomy paradox identified in research on healthcare AI deployment, where systems described as agentic in commercial contexts operate with heavily constrained autonomy in deployment, and accountability becomes untethered from actual performance when governance frameworks respond to promissory capabilities rather than to deployed behavior (Dias et al., 2026). The recovery mechanism and its accompanying responsibility architecture tether accountability to the actual moment of human decision rather than to the nominal presence of a human in the system loop.

## 5. Cross-Domain Implications

### Healthcare and Clinical Decision Making

In clinical environments, the silent cost of AI assistance manifests as erosion of diagnostic reasoning capacity in clinicians who rely extensively on AI-generated recommendations (Budzyń et al., 2025). The pattern identified in HIAG, in which AI delegation occurs with formal human oversight that does not constitute genuine evaluation, precisely describes the condition in which the surrender threshold is approached without awareness (Margondai et al., 2025). A recovery mechanism in clinical AI design would require not merely that a physician review an AI recommendation, but that the design architecture supports restoring active clinical reasoning prior to that review. This represents a fundamental challenge to the current model of AI-assisted clinical decision support, which typically presents AI outputs in formats that encourage acceptance rather than evaluation (Lyell & Coiera, 2017).

### Defense and Human-Autonomy Teaming

In defense contexts, the consequences of approaching the surrender threshold are potentially irreversible on operational timescales. Human-autonomy teaming research has established that out-of-the-loop performance decrements are a persistent challenge in systems where autonomous agents manage high-tempo information environments and human operators maintain supervisory roles (Endsley & Kiris, 1995; Parasuraman et al., 2000; O'Neill et al., 2022). The present model extends this concern to the cumulative dimension: operators who have spent extended periods in supervisory rather than active roles may find that their capacity for autonomous function in degraded or denied communications environments has eroded beneath the threshold necessary for effective performance. The recovery mechanism in defense-relevant AI systems must therefore incorporate active maintenance of human autonomous function as a standing design requirement rather than an emergency provision.

### Everyday Information Environments

The broadest domain of application for the present model is the everyday information environment in which billions of individuals engage with AI-mediated recommendation, navigation, search, and communication systems. In these environments, the silent cost of autonomy surrender is distributed across the full range of human cognitive and social functioning. Recommendation algorithms reduce the exercise of personal taste formation; navigation systems reduce spatial reasoning (Dahmani & Bohbot, 2020); search AI reduces the evaluation of information; conversational AI reduces the cognitive demand of self-expression and argumentation. No single transaction constitutes a significant surrender, and the cumulative effect builds over years of use, with no moment at which the cost becomes legible to the individual.

The model predicts that individuals in these environments will exhibit asymmetric difficulty with re-entry: they will find autonomous function in AI-mediated domains more cognitively demanding over time, will experience increased uncertainty and discomfort in the absence of AI guidance, and will, in advanced cases, exhibit a functional dependence that is experienced not as a deficit but as a preference, a terminal state the present model terms preference inversion. Preference inversion must be distinguished from rational delegation, in which an individual prefers AI assistance for reasons of comparative advantage while either retaining latent unaided capacity or accurately recognizing what has been traded away. The construct applies only when three conditions jointly hold: the individual prefers continued delegation, can no longer perform the function unaided at the criterion level, and discounts or denies the loss. Absent the capacity and awareness criteria, a preference for assistance is efficiency, not pathology. Preference inversion is particularly important for understanding the social and political dimensions of AI governance: a population whose autonomous cognitive function has been substantially replaced by AI-mediated processing may not experience the desire to reclaim that function, rendering the recovery mechanism not only a design challenge but a cultural and political one.

## 6. Discussion

### Relationship to Existing Frameworks

The present model builds most directly on the HIAG framework and the Transparency Paradox model (Margondai et al., 2025), while extending toward chronic rather than acute effects and cumulative rather than transient costs. Relative to the cognitive offloading literature (Risko & Gilbert, 2016), the present model contributes the concept of the surrender threshold and the re-entry barrier as distinct theoretical constructs that explain why offloading is not a fully reversible transaction. Relative to MHC and variable autonomy frameworks (Santoni de Sio & van den Hoven, 2018; Methnani et al., 2021), the model introduces the bandwidth restoration requirement as a necessary condition for meaningful control, which is absent from existing governance architectures.

The model is also in productive tension with the enrichment paradox literature, which describes similar phenomena of irreversible AI dependency but does not develop the re-entry architecture or the responsibility distribution that follows from it (Park et al., 2026). The present paper proposes that the irreversibility identified in that literature is threshold-dependent rather than absolute and that the design of recovery mechanisms oriented toward bandwidth restoration represents a tractable engineering and governance response to the challenge.

### Theoretical Implications

The first theoretical implication of the present model is a reconceptualization of autonomy itself. Psychological theory has predominantly treated autonomy as a trait-like disposition or a need to be satisfied, as in Self-Determination Theory (Deci & Ryan, 2000), while governance theory has treated it as a status to be respected. The present model treats autonomy as a dynamic, depletable, and restorable resource with measurable state dynamics and slowly varying structural capacity. This shift has consequences beyond human-AI interaction: any theory that invokes autonomous choice as an explanatory primitive inherits an obligation to specify the conditions under which the capacity for that choice is available, and the two-timescale architecture proposed here offers a template for that specification.

Second, the model unifies what have been treated as separate literatures under a single causal architecture. The acute, within-session effects documented in transparency and cognitive load research, the chronic deskilling effects documented in automation research, and the dependency phenomena emerging in the cognitive offloading literature have developed largely in isolation. The state bandwidth and structural capacity distinction connects them: acute saturation is the mechanism by which each transaction extracts its silent cost, and accumulated extraction is the mechanism by which structural capacity erodes. The model thereby converts a set of parallel empirical observations into stages of a single process, which is the characteristic contribution of a framework rather than a finding.

Third, the surrender threshold converts the deskilling literature from descriptive to predictive science. Existing accounts establish that disuse degrades skill; the threshold construct predicts the functional form of that degradation, specifies the individual-level analysis required to detect it, and identifies the quantity that intervention design must target. A literature that can state where the cliff is located is categorically more useful than one that can only report that people sometimes fall.

Fourth, the conditional responsibility architecture contributes to the theory of accountability in sociotechnical systems. The persistent difficulty in AI ethics has been that attributions of responsibility detach from causal capacity: the human in the loop absorbs blame for systems they could not realistically govern (Elish, 2019). By gating responsibility transfer on demonstrated restoration of capacity, the model supplies the missing theoretical link between having authority and being accountable, and it does so in a form that is, in principle, auditable rather than merely declaratory.

**Practical Implications**

For system designers, the model implies that re-entry support is a functional requirement, not an ethical garnish. Concretely, this means interfaces that incorporate prediction-first interaction patterns in evaluative domains; pacing architectures that distinguish informational saturation from informational need rather than maximizing throughput; explicit recovery affordances that terminate output, structure a period of cognitive quiet, and stage the return of decisional authority; and instrumentation that logs the human's independent performance over time so that approach to the surrender threshold is visible to the system, the user, and the organization. Design teams should treat a user's deliberate disengagement as a success metric, which inverts the engagement-maximizing logic that currently governs commercial development.

For organizations that deploy AI in high-consequence domains, the model implies that human capability maintenance must be managed as actively as system performance. In clinical settings, this argues for scheduled unaided practice and for decision-support configurations that require committed judgment before recommendations are displayed. In defense and aerospace settings, where degraded or denied environments make autonomous human function a contingency requirement rather than a preference, it argues for standing proficiency rotations in which operators perform without algorithmic support, and for treating measured unaided performance as a readiness indicator alongside equipment status. The cost of such maintenance is real; the model's claim is that it is the price of keeping meaningful human control meaningful.

For regulators and standards bodies, the model supplies the operational content that existing oversight mandates lack. Requirements for human oversight, as in the EU AI Act's provisions for high-risk systems, currently specify the presence of a human without specifying the conditions of that human's competence. The framework implies three auditable requirements: that systems classified as high-risk demonstrate a re-entry pathway in design documentation; that deployers monitor and report indicators of operator capacity over time rather than at certification alone; and that responsibility allocation in post-incident analysis examines whether the conditions for genuine re-entry were met, rather than defaulting to the most proximate human. Oversight that ignores capacity produces compliance theater; oversight that measures it produces the accountability the mandates intend.

Finally, for individuals, the model implies that autonomy preservation is a practice rather than a possession. Deliberate periods of unassisted performance in domains a person considers identity-relevant, attention to the subjective markers of the re-entry barrier, such as growing discomfort with unaided judgment, and skepticism toward one's own preferences for delegation in domains where unaided capacity has not recently been tested are the individual-level analogs of the recovery mechanism. The model does not counsel abstention from AI assistance, which would forfeit its genuine benefits; it counsels the same discipline toward cognitive capacity that is uncontroversial for physical capacity: namely, that what one wishes to retain, one must continue to use.

**Limitations and Future Directions**

The present paper is theoretical in character, and the model proposed here requires empirical validation across several of its central constructs. The surrender threshold requires operationalization through measurable behavioral and physiological markers. Prior work by the first author on ECG-derived heart rate variability as a precursor signal for cognitive state transitions (Margondai, in preparation) suggests a promising avenue: if the approach to the surrender threshold produces detectable physiological signatures analogous to the $W_2$ Wasserstein precursor signals identified in the IGMT framework developed in the first author's dissertation research, then the threshold may be detectable in advance of its functional consequences and thereby addressable before re-entry becomes difficult.

The model's central distinguishing claim is falsifiable and should be stated as such. That disused skills decay is established science; what the present model adds is the prediction that re-entry difficulty is nonlinear in cumulative delegation. A longitudinal design comparing matched AI-assisted and unassisted groups, with periodic forced unaided performance probes, discriminates between the competing accounts. Pure skill-decay models predict gradual, approximately linear degradation of unaided performance as a function of delegation history. The present model predicts a threshold-shaped, sigmoidal trajectory in which unaided performance is relatively preserved up to a critical accumulation of delegation and deteriorates sharply beyond it, accompanied by disproportionate increases in re-entry latency, intolerance of decisional uncertainty, and reluctance to disengage from assistance. Observation of strictly linear decay across the full range of delegation histories would falsify the surrender threshold construct; observation of the predicted discontinuity would corroborate it and would simultaneously identify the empirical location of the threshold that recovery mechanism design must respect. One methodological requirement is non-negotiable: because the threshold location varies across individuals, a mixture of individually sigmoidal trajectories with heterogeneous inflection points can aggregate to a smooth, approximately linear group curve, so group-mean analyses can falsely reject the model even when every individual conforms to it. The discriminating test must therefore be conducted at the individual level, using mixed-effects growth models or change-point analyses that can detect person-specific discontinuities.

Bandwidth restoration requires empirical characterization of its temporal dynamics, individual variability, and its relationship to domain expertise, cognitive reserve, and prior AI exposure. The responsibility architecture of the recovery mechanism raises normative and legal questions that extend beyond the scope of this paper and that will require engagement with AI governance scholars, legal theorists, and ethicists. The cultural and political dimensions of the everyday information environment application represent a research agenda in themselves.

### 7. Conclusion

Artificial intelligence systems extract a silent cost from every human who uses them. That cost is the incremental surrender of autonomous cognitive function in exchange for the products of AI processing. The exchange is structurally asymmetric, experientially invisible, and cumulative over time. When the accumulated surrender approaches the threshold at which re-entry becomes cognitively and psychologically difficult, the design imperative of meaningful human control can no longer be satisfied by merely providing an override option. It requires the active restoration of the cognitive bandwidth that makes control meaningful.

The recovery mechanism proposed in this paper is not a limitation on AI capability. It is a recognition that the human in the loop is not simply a compliance requirement but a cognitive entity whose capacity for autonomous function must be actively maintained if it is to be genuinely available. The AI system that delivers information and then creates space for the human to process, decide, and own the outcome is not a lesser system. It is a more honest one.

The ethics of AI design have focused extensively on what AI systems must not do: they must not deceive, discriminate, or exceed the scope of human authorization. The present paper advances a positive obligation: AI systems must preserve the conditions for the human autonomy they are trusted to support. That obligation begins with acknowledging the silent cost and ends with designing the quiet that makes reclamation possible.